# Personalized Auto-Grading and Feedback System for Constructive Geometry Tasks Using Large Language Models on an Online Math Platform

Yong Oh Lee[*]     Byeonghun Bang     Joohyun Lee     Sejun Oh[†]

September 29, 2025

**Abstract**

As personalized learning gains increasing attention in mathematics education, there is a growing demand for intelligent systems that can assess complex student responses and provide individualized feedback in real time. In this study, we present a personalized auto-grading and feedback system for constructive geometry tasks, developed using large language models (LLMs) and deployed on the Algeomath platform, a Korean online tool designed for interactive geometric constructions.

The proposed system evaluates student-submitted geometric constructions by analyzing their procedural accuracy and conceptual understanding. It employs a prompt-based grading mechanism using GPT-4, where student answers and model solutions are compared through a few-shot learning approach. Feedback is generated based on teacher-authored examples built from anticipated student responses, and it dynamically adapts to the student's problem-solving history, allowing up to four iterative attempts per question.

The system was piloted with 79 middle-school students, where LLM-generated grades and feedback were benchmarked against teacher judgments. Grading closely aligned with teachers, and feedback helped many students revise errors and complete multi-step geometry tasks. While short-term corrections were frequent, longer-term transfer effects were less clear. Overall, the study highlights the potential of LLMs to support scalable, teacher-aligned formative assessment in mathematics, while pointing to improvements needed in terminology handling and feedback design.

# 1 Introduction

The growing demand for personalized learning in mathematics education has been increasingly supported by the integration of online learning platforms with intelligent systems capable of delivering individualized feedback [1, 2]. The advent of the digital era, accelerated by the COVID-19 pandemic, has popularized remote education and driven the rapid expansion and familiarity of diverse online platforms among students [3]. Younger generations' early exposure to information and communication technologies has further stimulated the development of interactive educational tools. To realize the full potential of online mathematics education, these platforms must evolve beyond merely digitizing classroom instruction and instead leverage digital features to provide interactive, context-aware, and personalized learning experiences, particularly critical in cognitively demanding domains such as mathematics [4, 5].

---

[*]Hongik University, email: yongoh.lee@hongik.ac.kr
[†]email: soh@hongik.ac.kr

Despite this expansion, a significant gap remains in offering timely, specific, and individualized feedback for complex, open-ended, inquiry-based, or constructive problem-solving tasks, where students use digital tools to explore and creatively solve real-world problems. Most existing systems predominantly depend on answer-based grading, limiting their ability to capture the full spectrum of student reasoning and provide pedagogically rich feedback [4]. Additionally, online environments typically engage learners at diverse levels, posing challenges for educators to deliver extensive personalized feedback, especially in large classes with heterogeneous learners. Traditional classroom settings further constrain teachers' capacity to fully address individual learning trajectories and needs.

This study addresses this gap by focusing on constructive geometry tasks(CGT), which require students to perform multi-step visual constructions using logical and spatial reasoning. Unlike traditional closed-ended tasks, constructive geometry tasks encourage a variety of creative solution strategies. For example, students may determine a fair meeting point by calculating the circumcenter—the point equidistant from three given coordinates—or compute the radius of a broken circular artifact using coordinate-based geometric modeling. Rooted in real-world and inquiry-driven contexts, these tasks demand not only procedural accuracy but also deep conceptual understanding. To succeed, students employ educational engineering tools within the online platform that support translating practical problems into coordinate systems and facilitate geometric reasoning. These tools enable systematic solution structuring, justification of methodological decisions, and drawing valid conclusions, creating an effective environment for personalized formative assessment.

To this end, we developed a personalized auto-grading and feedback system that integrates a large language model (e.g., generative pre-trained transformer, GPT) with Algeomath[6], a Korean web-based platform for interactive geometry tasks. The system is designed to (1) automatically grade diverse student responses by recognizing varied problem-solving approaches, (2) provide adaptive, context-aware feedback referencing students' previous attempts across up to four iterations, and (3) investigate the educational effectiveness of combining a dynamic geometry tool with a language model–based evaluation engine.

Rather than relying on static rubrics, the system leverages teacher-authored feedback examples built from anticipated student responses, enabling the language model to generate targeted and pedagogically sound guidance. This approach seeks to bridge the gap between scalable automation and the quality of human-level feedback.

The contributions of this study are threefold: (1) the design and implementation of a system specifically tailored for online constructive geometry education, (2) the application of few-shot learning prompts to evaluate the adaptability and accuracy of LLM-based grading and feedback, and (3) the collection and analysis of real classroom data from 90 students to assess the system's practical effectiveness and educational value.

## 1.1 Related Works

### 1.1.1 Personalized Learning and Intelligent Tutoring Systems

Early work on automated assessment in mathematics focused on evaluating students' free-form answers (textual or mathematical) in online platforms. Erickson et al. demonstrated the feasibility of grading over 150,000 open-ended student responses in ASSISTments, which is a popular online math tutor(https://www.assistments.org/) using machine learning models [7]. They compared classifiers like decision trees, random forests, and long short-term memory(LSTM) for scoring short answers containing both text and equations, finding that LSTM can approximate human graders on multi-point rubrics [7, 8].

Wu et al. designed a dynamic adaptive testing system for fraction arithmetic that gave immediately individualized feedback on each problem [9]. If a student answered incorrectly, the system would diagnose the misconception (e.g. confusion between numerator and denominator roles) and provide a tailored hint targeting that error. As the student continued, the system adjusted problem difficulty and the specificity of feedback (more detailed hints after repeated mistakes). In a study with 118 students, they found that those who received adaptive tests with just-in-time feedback outperformed students who received static practice or only end-of-chapter reviews [9]. This underscores that instant, adaptive feedback can substantially boost understanding in math learning.

Another study is incorporating affective computing into personalized learning. Learning is not only cognitive but also emotional – frustration or boredom can impede progress. Grawemeyer et al. tackled this by making an ITS affect-aware [10]. In their iTalk2Learn system for fractions, the tutor used students' speech and interaction patterns to infer emotional states like engagement, confusion, or boredom. They developed Bayesian models to predict which feedback strategy would most likely keep each student motivated. For instance, if a student showed signs of frustration, the system might provide motivational feedback or hints to bolster confidence, whereas for a bored student it might pose a reflective question to re-engage them. In classroom evaluations, the affect-aware version of the tutor significantly reduced students' off-task behavior and boredom, compared to a version that ignored affect [10]. This led to improved persistence and better learning outcomes, as the tutor could respond to disengagement before learning stalled. These results highlight that personalization in ITS can extend beyond content difficulty to the style and timing of feedback, tuned to students' emotional needs.

### 1.1.2 Large Language Models in Mathematics Education

The advent of powerful Large Language Models (LLMs) like GPT-3 and GPT-4 has catalyzed new approaches to assessment and tutoring in math. Unlike rule-based tutors, LLMs can parse and generate natural language, reason through complex problems, and engage in dialogue – capabilities that can be leveraged for both automated grading and conversational tutoring. Researchers have begun using LLMs not just to solve problems, but to judge student-written solutions. Gao et al. introduced Omni-Judge, a math-specific evaluator model built on LLaMA and fine-tuned with high-quality GPT-4 outputs [10, 11]. Omni-Judge was trained to decide if a solution is correct given the problem and a reference answer. It achieved remarkable consistency with expert grading – about 95% agreement with GPT-4's own evaluations [11]. Such LLM-based graders can assess not only final answers but also the validity of reasoning steps. Xia et al. similarly fine-tuned GPT-4 on a large dataset of step-by-step solutions (PRM800K) to perform reasoning evaluation, labeling each step as correct or not [12]. This approach focuses on the process, predicting errors at the step level rather than just the end result. In essence, these works treat LLMs as an automatic math TA, comparing a student's reasoning to known correct reasoning and flagging discrepancies.

One of LLMs' strengths is their ability to learn from examples at inference time (few-shot learning). This has been harnessed to generate solutions and hints. Ahn et al. showed that carefully selecting a few exemplars (with "chain-of-thought" reasoning) in the prompt can guide GPT-3/4 to produce better step-by-step solutions. They introduced PromptPG, a method to reinforce learning good prompts via feedback, which reduced the instability of few-shot performance on complex problems [13]. In educational settings, a teacher or system might prime an LLM with two well-explained sample solutions, which increases the likelihood that the model will evaluate or support a new solution correctly. Another technique, Progressive Hint Prompting, involves asking an LLM to first offer a small hint, then a larger one, before finally revealing the answer, thereby simulating scaffolding [14]. Studies report that this can boost correctness in certain subjects (e.g., GPT-4 on abstract algebra problems), although the effect is not uniform. In general, prompt engineering has become an important strategy for making LLMs educationally useful rather than just accurate.

LLMs have shown a remarkable ability to generate natural language feedback and hints. Unlike earlier systems that used fixed hint templates, an LLM can tailor a hint to the specific student answer. For example, Gupta et al. integrated an LLM into an ITS as a hint generator. The ITS provided the model with information about the student's current step and knowledge state, and the LLM then produced a contextual hint or error explanation [15, 16]. Teachers in their study noted that the LLM's hints were often detailed and mimicked a tutor's tone, sometimes even preferring them to generic feedback. LLMs can also rephrase explanations in simpler terms, a boon for students who need concepts in plain language. Khan Academy's experimental Khanmigo tutor is one real-world example – using GPT-4, it engages students in Socratic dialogue, giving step-by-step prompts rather than the answer outright. Similarly, Quizlet's Q-Chat uses an LLM to adjust question difficulty on the fly and provide guidance if a student is stuck [15]. These applications illustrate how LLMs enable a new level of interactive tutoring: they can converse with a student, understand free-form questions ("why did I get this wrong?"), and give constructive, personalized feedback in real time.

Despite these advantages, researchers caution about LLM hallucinations and oversight. LLMs sometimes introduce misconceptions – for example, explaining a concept in a technically wrong way – with a confident tone. Gupta et al. observed that only about 56% of the tutorial interactions generated by ChatGPT in their math tutoring prototype were fully correct on all points [16, 17]. In the remainder, the LLM made minor errors or gave overly complex explanations that could confuse learners. This aligns with [10, 18], which emphasizes that human teachers must stay in the loop to vet and modify AI-generated answers. Teachers need to be aware of the AI's limitations – such as occasional factual errors or lack of pedagogical tact – and intervene accordingly. Encouragingly, studies suggest that when teachers and AI work collaboratively (with teachers guiding the AI's use and double-checking its outputs), students benefit the most [9]. The AI can handle routine responses and provide diverse solutions, while the teacher ensures accuracy and appropriateness.

### 1.1.3 Digital Tools in Geometry Education

Research in math education has long recognized the value of dynamic and visual tools, particularly in geometry where visualization is key. Over the past decade, software like GeoGebra has become a staple for teaching geometry, and studies have examined how such tools influence learning and pedagogy.

Jablonski and Ludwig provided a comprehensive review of current trends in geometry education [19]. They identified major themes, one of which is the use of digital tools for geometric exploration. Dynamic Geometry Software (DGS) allows students to construct and manipulate figures, observing properties as they drag points. This interactivity aligns with theories of embodied cognition and can deepen understanding of invariants (properties that remain unchanged under transformations). The review noted that while many research findings are positive about DGS, a gap exists in transferring these insights to everyday classroom practice. Teachers often need guidance on integrating tools like GeoGebra into lessons in a pedagogically sound way.

A systematic review by Yohannes and Chen specifically examined GeoGebra integration from 2010–2020. They found that the majority of studies were done in high school geometry and calculus, often using task-based learning strategies where students discover concepts through guided activities in GeoGebra [20, 21]. Importantly, students generally showed improved conceptual understanding and higher-order thinking (e.g. conjecturing, problem solving) when using GeoGebra, compared to traditional methods. However, the review also pointed out areas under-researched, such as cognitive load and anxiety. Very few studies measured whether using dynamic software might overwhelm some students or how to scaffold its use for those who struggle. Another notable point was that student engagement is not automatic – while some are motivated by interactive visuals, others may get frustrated without proper prompts or guidance. Thus, the teacher's role remains critical in orchestrating tool-based activities.

Some experimental studies reinforce these findings. Birgin and Yazıcı conducted a quasi-experiment with 8th graders on linear equations and slopes, comparing a GeoGebra-supported lesson to a traditional one. The GeoGebra group showed significantly better conceptual understanding on post-tests and retained that understanding weeks later better than the control group [20, 22]. This was attributed to students visually exploring the effect of changing a line's slope in real time, which helped them link the abstract equation form $y = mx + b$ to the graphical meaning of m. Meanwhile, Ulusoy and Turuş analyzed Turkish math textbooks and found that while high school books included more DGS-based tasks than middle school books, many tasks were low-level (e.g. "use the software to draw this shape following steps") [20, 23]. Truly exploratory tasks – where students might discover a theorem or test a conjecture by dragging points – were relatively rare. The authors recommended enriching textbooks with deeper DGS tasks that require students to notice invariant properties (e.g. the sum of angles in a triangle stays 180° no matter how you reshape it) and make conjectures. This aligns with Nagar et al.'s work, who observed that teachers themselves often notice geometric invariants only when prompted in a DGS environment [20, 24]. In their study, high school teachers using a dynamic geometry program initially missed many invariant relationships until researchers explicitly asked them to look for what stayed constant as they dragged points. This suggests professional development is needed for teachers to fully utilize dynamic tools' potential – teachers must learn to "see" the math through the software and guide students to do the same.

Table 1: Structure and key features of each problem stage in the Algeomath task sequence.

| Stage | Description | Construction Tools | Assessment Type |
|---|---|---|---|
| 1 | Introducing the concept of triangle circumcenter and circumcircle | Perpendicular bisector, intersection | Closed-ended answer + Open-ended explanation |
| 2 | Constructing and exploring the circumcenter location | Segment, length, move tool | CGT – Conceptual + Open-ended explanation |
| 3 | Applied task using circumcenter to determine a fair meeting point | Circumcenter, measurement tool | CGT – Applied + Open-ended explanation |
| 4 | Justifying the result of Stage 3 with written explanation | Written text | Open-ended explanation |
| 5 | Investigating effect of triangle type (acute, obtuse) on circumcenter position | Angle, circumcenter, move tool | CGT – Conceptual |
| 6 | Reconstructing a traditional tile pattern using circumcenter geometry | Polygon construction, circumcenter | CGT – Applied |

## 2  Methods: Development of Auto-Grading and Feedback Systems

### 2.1  Mathematics Problem Design and Implementation on Algeomath

To systematically investigate the educational effectiveness of LLM-based auto-grading and feedback, a sequence of constructive geometry tasks was developed and implemented using Algeomath, a web-based dynamic geometry platform. The problem set was designed to foster both procedural accuracy and creative application in the context of the circumcenter of a triangle, culminating in a practical restoration scenario inspired by cultural heritage artifacts.

More detailed, the learning objectives of this problem set emphasize a comprehensive understanding and justification of the properties of the triangle's circumcenter and incenter. Students are expected to engage in observational and experimental activities using engineering tools, providing explanations supported by empirical evidence, analogical reasoning, and formal proof where appropriate to their proficiency level. Furthermore, the tasks encourage learners to apply these geometric concepts to real-life problem scenarios, fostering an appreciation of the relevance and value of mathematics and promoting reasoned decision-making based on mathematical insight.

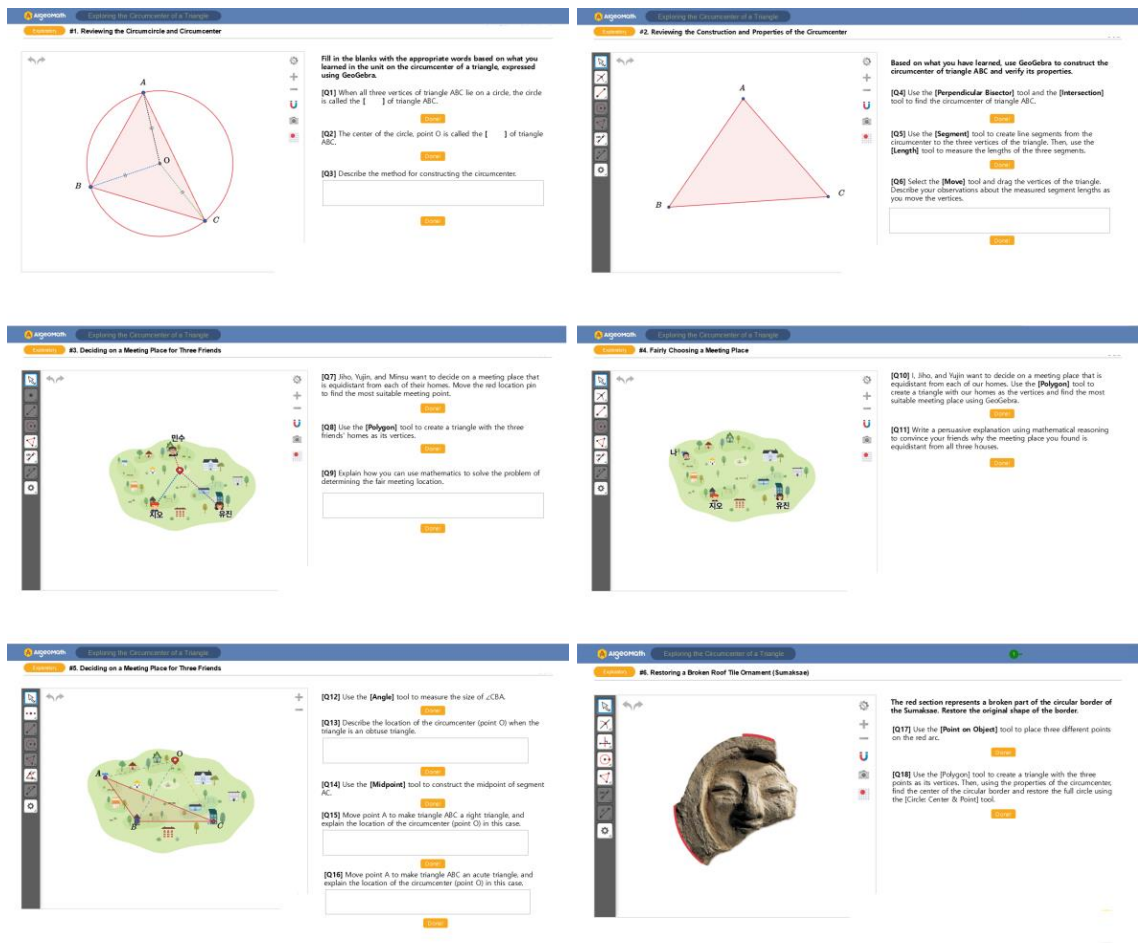

Figure 1: Screenshots for each stage of the Algeomath task sequence, visually demonstrating the progression from basic circumcenter construction (Stage 1–2), to applied problem-solving using the circumcenter in social and cultural contexts (Stage 3–6)

As depicted in Table 1, the problem set incrementally introduces and reinforces geometric concepts and skills. Students progress through short-answer recognition, hands-on construction and measurement, real-world application, mathematical explanation, exploration under varying geometric conditions, and algorithmic restoration of a cultural artifact.

Figure 1 displays screenshots for each stage of the Algeomath task sequence. Each screen is divided into two panels: the left panel embeds an interactive Algeomath environment where students engage in constructive geometry tasks (CGTs), while the right panel presents problem prompts and input boxes for students' responses. After completing a task—whether by submitting a closed-ended answer, constructing a geometric object, or writing an open-ended explanation—students can submit their responses. A turtle icon in the bottom-right corner then provides automated feedback and evaluation based on the submitted answer.

Throughout all stages, students interact directly with Algeomath's interface to create, measure, and manipulate geometric objects according to the problem requirements. Each response is automatically evaluated, and students receive adaptive formative feedback, with up to three retries per item permitted. This integration of hands-on digital construction, context-rich prompts, and automated guidance forms a comprehensive framework for analyzing both process and understanding in constructive geometry.

The design of this multi-stage problem sequence enables the system to capture a wide range of student reasoning processes and supports rigorous evaluation of LLM-based grading and feedback mechanisms.

## 2.2 Auto-Grading Using LLM

Following student submissions through the task interface, we developed a hybrid auto-grading system that leverages both deterministic algorithms and large language model (LLM) inferences, tailored to the distinctive requirements of constructive geometry tasks. The system is designed to ensure grading reliability, scalability, and the ability to accommodate the diversity of student responses across core assessment types.

For closed-ended answer cases, which typically ask students to identify specific mathematical terms, definitions, or properties, we choose a rule-based grading that is when a student submits a response, the server-side system directly compares the submission against a predefined set of correct answers. Given the objective and constrained nature of these items, fixed matching rules ensure rapid and unambiguous grading, entirely independent of LLM output variability or latency. This approach prioritizes grading speed, transparency, and robustness, which are essential for real-time feedback in high-frequency recall contexts.

For open-ended explanation cases, where students provide open-ended, written answers involving explanation, reasoning or justifications, the auto-grading system utilizes a few-shot prompting approach with an LLM. This method enables flexible evaluation of diverse student language expressions while adhering to pedagogical grading criteria.

At the point of answer submission, the system assembles a prompt for the LLM that includes the student response, the problem statement, and a curated set of exemplar model answers previously provided by math education experts. These exemplary answers serve as the reference standard indicating acceptable correct approaches.

The LLM is tasked with producing a structured output that indicates the correctness of the student's answer. Specifically, the output follows a predefined schema—implemented, for instance, via a Python data model with fields such as a boolean correctness flag. This schema-driven design facilitates reliable parsing and downstream utilization of the grading result.

Using few-shot examples embedded within the prompt, the LLM learns to classify the student submission as correct if it sufficiently incorporates the conceptual elements present in at least one model answer. This approach balances the ability to generalize beyond the exact wording of provided examples with the need for consistency and alignment with teacher expectations.

The grading pipeline leverages prompt template frameworks that is *FewShotChatMessagePromptTemplate* of LangChain [25] to construct the conversational context and examples. Upon receiving the LLM's structured evaluation, the system extracts the correctness judgment and then the correcteness judgement is aligned with feedback generation introduced in the next subsection.

Using LLM's capability with few-shot prompts enables automated, scalable, and pedagogically valid grading of complex, natural language mathematical explanations, accommodating a range of valid student expression styles in open-ended explanation while maintaining adherence to defined instructional standards.

For CGT cases, where students use the dynamic geometry platform (Algeomath) to perform constructions and measurements, automated grading is accomplished through a combination of programmatic object extraction and rule-based validation. When a student submits a solution, the platform transmits a serialized representation of their construction session to the server. This string encodes all objects, measurements, and their relationships within the workspace.

More specifically, the system parses the submitted data to extract relevant geometric objects. Key geometric objects, summarized in Table 2 implemented and utilized within Algeomath for construction and grading purposes include points, lines and segments (including perpendicular bisectors and midpoints), circles (including circumcircles and incenters), polygons, angles, and measurement tools for distances and angles. These objects form the foundational elements for representing students' geometric solutions and enable precise, rule-based assessment grounded in the structural and quantitative properties of submitted constructions. These data are typically encapsulated within an "object capsule" data structure. For each item, the grading process follows a task-specific logic that inspects the geometric constructs for the required features.

Table 2: Key Geometric Objects Utilized in Algeomath for Inquiry-Based Tasks

| Category | Object/tool name | Application Example |
| --- | --- | --- |
| Point | Point, Intersection Point, Movable Point | Triangle vertices, intersection loci |
| Line/Segment | Line, Segment, Perpendicular Bisector, Midpoint | Triangle sides, bisector constructions |
| Circle | Circle, Circumcircle, Incenter Circle | Circumcenter constructions, loci |
| Polygon | Polygon | Triangle, roof tile boundary reconstruction |
| Angle | Angle, Angle Marker | Measuring angle sizes, verifying properties |
| Measurement | Length, Distance, MeasureAngle | Segment lengths and angle magnitudes |

The grader implements task-specific routines to verify the existence and correctness of these objects as required by each problem. For example, the presence and accuracy of a perpendicular bisector or a correct angle measurement are programmatically checked against expected geometric configurations and numerical tolerances. Only submissions satisfying these criteria are marked as correct.

This CGT grading approach provides robust, scalable evaluation that depend heavily on students' procedural constructions and measurement submissions within a dynamic geometry environment. Also, CGT grading ensures objective evaluation of diverse solution strategies in inquiry-based geometry tasks while maintaining scalability and efficiency suitable for real-time grading environments.

The integration of these distinct grading pipelines enables the system to address the full breadth of mathematical task types present in constructive geometry education. The design ensures that each assessment type is evaluated using methods appropriate to its input complexity and pedagogical intent, thereby supporting both fairness and authenticity in student evaluation. A summary table illustrating the procedural flow and key differences across grading types is presented at the end of this section.

## 2.3 Feedback generation using LLM

In our system, the primary role of LLM is not to determine grades for open-ended explanation, but to generate personalized and context-sensitive feedback based on students' submission history and performance. This feedback generation is grounded in expert-validated teacher exemplars and aims to guide students toward deeper understanding. Building upon the grading process described earlier, the feedback mechanism dynamically tailors responses according to correctness judgments and the number of remaining attempts for the given problem.

Teacher-authored feedback examples, anticipated in advance for each question and potential student solution paths, form the foundation of this approach. These exemplars encapsulate pedagogically sound hints, clarifications, and explanations designed to scaffold student learning at varying levels of detail.

In addition, the feedback generation pipeline incorporates the following principled strategies based on attempt count. In the early attempts, when students still have sufficient opportunities to revise their answers, the generated feedback emphasizes strategic guidance that encourages students to reflect on and recalibrate their problem-solving approaches. Feedback at this stage consists of hints or suggestions that draw attention to key aspects of the problem or common errors without explicitly revealing the solution. As the number of attempts decreases, the feedback shifts toward more direct corrective guidance. At this point, it specifically identifies common misconceptions or errors present in the student's response and clearly explains why these aspects are incorrect, helping students recognize and correct their misunderstandings. When the final attempt has been exhausted without arriving at a correct answer, the feedback transitions into a fully explanatory and instructive mode. In this phase, model answers accompanied by detailed rationales are provided, enabling students to thoroughly understand the correct solution process and the underlying concepts, thereby promoting productive learning even when errors persist.

For instance, consider Stage 5's task requiring students to measure the interior angle CBA using Algeomath's angle measurement tool. When a student correctly measures the interior angle CBA, the system acknowledges the correct response and provides positive feedback such as "Triangle ABC is an obtuse triangle." If the student fails to measure the angle or submits an incorrect measurement—for example, measuring the major (exterior) angle instead—the system responds with instructive feedback guiding the student to use the angle measurement tool properly, for instance: "Please use the angle measurement tool to measure the interior angle CBA," or "The angle you measured is not the interior angle. To measure the interior angle correctly, click the points in clockwise order (C $\rightarrow$ B $\rightarrow$ A)."

Upon incorrect first attempts, students receive a second chance, accompanied by progressively more explicit feedback tailored to encourage their reasoning in alignment with the learning objectives. For example, during the first retry, the feedback might gently remind the student: "Triangles can be classified into three types based on their interior angles. Recall what you have learned." If the student does not improve, the feedback becomes more descriptive: "The angle CBA is greater than 90 degrees. What type of triangle does that indicate?" On the final attempt, explicit clarification is given: "The angle CBA exceeds 90 degrees; therefore, triangle ABC is an obtuse triangle."

This staged feedback approach exemplifies how teacher-authored feedback templates can guide students toward conceptual understanding through adaptive, supportive prompts that scaffold learning effectively within an automated system.

Operationally, feedback generation leverages LLM capabilities via conditioned prompting. For each submitted response, the system assembles a prompt comprising the student's answer, the problem statement, relevant teacher-authored feedback examples, and the current attempt number. The LLM is then instructed to produce feedback aligned with the policies above, balancing pedagogical effectiveness with tone and length constraints suited for online learning environments.

In practice, the feedback prompt delivered to the LLM includes targeted behavioral instructions to ensure consistency, clarity, and pedagogical effectiveness across all feedback instances. The system imposes strict policies that permit up to three feedback opportunities per item, requiring the LLM to progressively tailor its responses according to the attempt count. During the first and second attempts, feedback must avoid revealing the correct answer; instead, it should provide constructive hints or explicitly address areas where the student's approach was incomplete or incorrect. For example, encouraging phrases such as "Would you like to try again by clicking the 'Retry' button?" are appended to first-attempt feedback to motivate students to make another effort. On the second attempt, the prompt always cues the system to begin with a supportive reminder—"Unfortunately, that wasn't correct, but don't give up yet!"—before delivering more specific guidance. For the final, third attempt, feedback opens with a clear indication that no further retries are available, using statements like "This is your final chance!" before providing an explicit explanation or the correct solution.

To maintain engagement and reinforce positive behaviors, the prompts are also designed to deliver personalized encouragement such as "So close!", or "Good attempt!" for answers that are close or show significant effort. Particularly for high-difficulty tasks where students succeed, additional messages such as "Please help classmates who may be having difficulty." may be provided to further encourage collaborative learning.

Strict linguistic style guidelines are applied to ensure that all feedback is presented in a polite, clear, and encouraging manner. Requests to students are formulated using soft, courteous expressions such as "Could you please…" or "Please try to…," promoting engagement without sounding overly directive. For example, instead of a terse "Please measure the angle," feedback might say, "Could you please measure the angle?" or "Please try measuring the angle."

Explanatory statements are written as complete, natural sentences that clearly convey the intended meaning. Rather than a literal "It is an obtuse triangle," the phrasing uses straightforward declarative sentences such as "Triangle ABC is an obtuse triangle," or "This means that triangle ABC has an obtuse angle." These forms provide clarity while maintaining an academic tone appropriate for educational feedback.

By adopting these linguistic conventions, the system ensures that feedback not only remains accessible and motivating to learners but also aligns well with professional standards for clear and supportive educational communication.

Each response is constrained to no more than 100 characters (preferably around 60) to ensure clarity and accessibility. Mathematical expressions are rendered using LaTeX formatting to enhance readability and align with best practices for communicating mathematical content in digital environments. The chatbot exclusively outputs feedback, omitting any extraneous commentary, and mathematical formulas are presented in proper LaTeX style to prevent confusion from raw code or markup.

The integration of structured, expert-curated feedback exemplars and attempt-aware dynamic prompting allows the system to approximate human-like instructional support, combining scalability with responsiveness to individual learner trajectories. This design complements the earlier automated grading framework, creating a closed-loop formative assessment environment that promotes iterative learning and sustained engagement.

# 3 Results: Field Study and Its Analysis

## 3.1 Overview of the Field Study

The field study was conducted in November 2024 at four middle school classes in South Korea. A total of 79 students participated in the study, engaging with a series of inquiry-based mathematics problems introduced in the previous section.

For each student submission, the system recorded key attributes including an anonymized student identifier (student_id), the problem identifier (question_id), the attempt number for that problem (attempts), and the correctness status of the response (answer_status (either correct or incorrect). The submitted content was stored as a unified record encompassing both the student's answer (answer) and any associated feedback (feedback).

For analysis, the stored data was refined to address several technical issues observed during the field sessions. In some cases, students' attempt counts exceeded the maximum of four due to repeated transmissions from unstable network connections. There were also instances where students continued to submit answers to the same question even after receiving a CORRECT judgment, often caused by interface delays or accidental resubmissions. In addition, intermittent database write errors occasionally led to multiple entries being recorded with the same attempt number for the same student and question, creating duplicated submission records. To ensure data integrity, all such anomalous entries were removed, and only records that satisfied these criteria were retained for the analyses described in the following subsections.

## 3.2 Evaluation of Grading Accuracy and Feedback Appropriateness

To begin, the grading results were examined in terms of their agreement with teacher assessments. In total, the dataset consisted of 18 questions, including two closed-ended answer cases (Q1, Q2), nine CGT cases (Q4, Q5, Q7, Q8, Q10, Q12, Q14, Q17, Q18), and seven open-ended explanation cases (Q3, Q6, Q9, Q11, Q13, Q15, Q16).

The first assessment concerned grading accuracy, namely whether student answers were classified as correct or incorrect. For closed-ended and CGT questions, the grading process was entirely rule-based; therefore, the comparison with teacher scoring yielded a perfect agreement rate of 100%.

For the open-ended explanatory tasks, grading was performed using an LLM rather than deterministic answer matching. Table 3 summarizes the performance across the seven explanatory items. Agreement rates varied by item, ranging from 71.3% (Q15) to 96.5% (Q13). Cohen's $\kappa$ values confirmed this trend, with some questions (e.g., Q13, $\kappa$ = 0.929) showing almost perfect agreement, while others such as Q15 ($\kappa$ = 0.434) revealed only moderate reliability. When aggregated across all explanatory tasks, the overall agreement rate reached 0.866 and the corresponding Cohen's $\kappa$ was 0.737, indicating substantial consistency between automated and teacher grading.

Table 3: Agreement rate and Cohen's $\kappa$ for open-ended explanatory questions

| Question | n | Agreement rate | Cohen's $\kappa$ |
|---|---|---|---|
| Q3 | 164 | 0.854 | 0.699 |
| Q6 | 130 | 0.900 | 0.802 |
| Q9 | 115 | 0.870 | 0.717 |
| Q11 | 116 | 0.828 | 0.629 |
| Q13 | 113 | 0.965 | 0.929 |
| Q15 | 136 | 0.713 | 0.434 |
| Q16 | 104 | 0.923 | 0.842 |
| **All** | **878** | **0.866** | **0.737** |

A more detailed breakdown revealed systematic differences in teacher agreement depending on whether the system judged the answer as correct or incorrect (Table 4). Across all explanatory questions, teachers almost always endorsed system judgments when the system labeled answers as correct (95.3% agreement). However, when the system labeled answers as incorrect, agreement dropped to 77.2%. This asymmetry suggests that while the LLM is highly reliable in recognizing correct solutions, its identification of incorrect responses occasionally conflicted with teacher interpretations, often due to partially correct reasoning or alternative valid expressions.

Table 4: Teacher agreement by system judgment (open-ended explanatory questions)

| System judgment | n | Teacher agreement (n) | Agreement rate |
|---|---|---|---|
| Correct | 426 | 406 | 0.953 |
| Incorrect | 452 | 349 | 0.772 |

An inspection of the cases where teachers disagreed with the system further illustrates these challenges. For example, in Q3 (finding the circumcenter of a triangle), the system often marked answers incorrect when students described the construction process with minor omissions or phrased the concept differently (e.g., mentioning "two perpendicular bisectors" instead of "three"). Teachers considered many of these responses acceptable, emphasizing conceptual understanding rather than exact wording. Similarly, in Q9 and Q11, discrepancies emerged when students used terms like "center" or "incenter" instead of the technically precise "circumcenter." Teachers frequently accepted these as valid, while the system penalized them. In Q15, the lowest-performing item, many disagreements arose from variations in describing the circumcenter of a right triangle (e.g., "on the hypotenuse" vs. "at the midpoint of the hypotenuse"), leading to misclassification by the system. These cases underscore that disagreements were not random but stemmed from semantic flexibility and mathematical register that human teachers were more willing to interpret generously.

Taken together, these analyses demonstrate that the LLM-based grading system achieves substantial agreement with teacher judgments in explanatory tasks, especially when recognizing correct solutions. However, further refinement is needed to better accommodate the diversity of mathematically valid expressions and partially correct reasoning that human teachers are able to recognize.

Following the system's correctness judgments, a mathematics education expert qualitatively evaluated the quality of feedback for those responses deemed correct. The evaluation employed a four-dimensional rubric encompassing accuracy, clarity, instructional guidance, and linguistic tone, with each dimension rated on a five-point Likert scale. This procedure enabled a structured appraisal of the pedagogical quality and communicative effectiveness of the feedback messages.

Table 5 reports the summary statistics of the overall guidance score by item. Aggregating across all seven explanatory questions yields a pooled mean of 3.98 ($n$ = 754) with a pooled standard deviation of 1.78, suggesting that the feedback was generally viewed as educationally serviceable but with substantial variability across items. At the item level, Q9 attained the highest average score (mean = 4.55, SD = 1.26), followed by Q11 and Q16, indicating that in tasks centered on spatial or geometric interpretation, the system's messages were often sufficiently accurate and helpful. By contrast, Q15 showed the lowest mean (mean = 3.33, SD = 2.03), together with the largest dispersion, pointing to inconsistent quality and frequent issues in how key mathematical ideas were communicated.

Table 5: Summary of feedback guidance scores by item (evaluated only when teachers agreed with the system's grading).

| Question | Count ($n$) | Mean | SD |
|---|---|---|---|
| Q3 | 140 | 3.421 | 1.949 |
| Q6 | 116 | 4.043 | 1.737 |
| Q9 | 100 | 4.550 | 1.258 |
| Q11 | 96 | 4.448 | 1.375 |
| Q13 | 109 | 4.064 | 1.847 |
| Q15 | 97 | 3.330 | 2.035 |
| Q16 | 96 | 4.219 | 1.656 |
| **ALL** | **754** | **3.980** | **1.779** |

A close reading of disagreement rationales and free-text comments reveals several recurring patterns. First, *model-answer leakage* was common: many feedback utterances effectively disclosed the solution (e.g., "the circumcenter is the midpoint of the hypotenuse"), which experts deemed pedagogically inappropriate for formative guidance. Second, *terminological imprecision* (e.g., using "center" instead of "midpoint" or conflating incen-ter/circumcenter) and *overly generic prompts* (e.g., "try again" without actionable next steps) depressed ratings of accuracy and clarity. Third, *tone escalation* such as "this is your last chance" undermined the supportive learning climate and was penalized in the tone dimension. In contrast, feedback that decomposed the reasoning into actionable steps (e.g., "draw two perpendicular bisectors and locate their intersection") and connected concepts explicitly received high marks, especially when it scaffolded the next attempt without revealing the answer.

Taken together, these findings indicate that the system generally produces pedagogically acceptable feedback when its grading aligns with teachers, but its effectiveness varies by item and by the specificity of the mathematical register. The analysis suggests concrete avenues for improvement: suppressing answer leakage, enforcing domain-precise terminology, and adopting principled scaffolding templates that promote stepwise reasoning without giving away the solution.

## 3.3 Quality and Usefulness of Feedback

Table 6: Number of students who answered correctly by attempt, final incorrect, non-attempt, and totals for each item

| Question | Type | 1st Correct | 2nd Correct | 3rd Correct | 4th Correct | Wrong All | Not Attempted | Total |
|---|---|---|---|---|---|---|---|---|
| Q1 | closed-ended | 43 | 22 | 6 | 4 | 4 | 0 | 79 |
| Q2 | closed-ended | 71 | 2 | 1 | 3 | 2 | 0 | 79 |
| Q3 | open-end | 30 | 12 | 10 | 3 | 24 | 0 | 79 |
| Q4 | CGT | 53 | 11 | 2 | 0 | 12 | 1 | 79 |
| Q5 | CGT | 36 | 15 | 6 | 3 | 13 | 6 | 79 |
| Q6 | open-end | 47 | 9 | 3 | 0 | 19 | 1 | 79 |
| Q7 | CGT | 37 | 12 | 2 | 1 | 25 | 2 | 79 |
| Q8 | CGT | 70 | 4 | 1 | 0 | 2 | 2 | 79 |
| Q9 | open-end | 55 | 10 | 3 | 0 | 10 | 1 | 79 |
| Q10 | CGT | 33 | 11 | 2 | 1 | 29 | 3 | 79 |
| Q11 | open-end | 50 | 9 | 8 | 1 | 9 | 2 | 79 |
| Q12 | CGT | 27 | 3 | 1 | 1 | 5 | 42 | 79 |
| Q13 | open-end | 46 | 7 | 4 | 1 | 18 | 3 | 79 |
| Q14 | CGT | 26 | 5 | 0 | 1 | 6 | 41 | 79 |
| Q15 | open-end | 31 | 13 | 5 | 2 | 25 | 3 | 79 |
| Q16 | open-end | 49 | 7 | 2 | 0 | 16 | 5 | 79 |
| Q17 | CGT | 36 | 17 | 6 | 0 | 14 | 6 | 79 |
| Q18 | CGT | 29 | 6 | 1 | 1 | 22 | 20 | 79 |

In this section, the educational value of the automated grading and feedback system is assessed through a comprehensive analysis of item-level evaluation metrics, learner performance trends, and qualitative patterns of response improvement. The study integrates psychometric indicators—such as question difficulty and discrimination indices—with learning behavior data, including attempt-level correctness and feedback-driven revisions, to derive pedagogically meaningful insights.

By synthesizing the results of question difficulty and discrimination analyses, learning curve modeling, and qualitative case studies of response refinement, the system's role in supporting geometry learning is evaluated from multiple perspectives

### 3.3.1 Difficulty and discrimination analysis

Difficulty indices were computed as the proportion of students who answered each question correctly on the *first* attempt, reflecting the relative ease or challenge posed by each task. Discrimination indices were operationalized as point–biserial correlations between the question's first–attempt score (0/1) and a total test score, with a leave–one–out (LOO) adjustment applied so that each question's contribution was removed from the total to avoid spuriously inflating the association. For questions with partial participation, the LOO discrimination was estimated on a completed student–by–question grid with non–attempts coded as 0, whereas the difficulty index uses the observed per–question denominator.

Figure 2 shows the difficulty indices. Across the 18 questions, first–attempt correctness ranged roughly from 0.42 to 0.91, with Q2 and Q8 exhibiting high ease (≈ 0.91) and several questions (e.g., Q3, Q10, Q15) presenting greater challenge (≈ 0.42–0.45). Figure 3 presents LOO discrimination, with most questions attaining moderate or higher values ($\gtrsim$ 0.30). The strongest discrimination was observed for Q10 (0.59), followed by Q4 (0.53), Q2 (0.54), and Q13 (0.51). Lower discrimination emerged for very easy questions such as Q8 (0.14) and relatively easy–to–mid questions such as Q7 (0.21).

The joint pattern is summarized in Figure 4. Questions in the mid–difficulty band (roughly 0.5–0.7) tended to discriminate best, whereas very easy questions (high difficulty index) showed attenuated discrimination, as expected. These results indicate that the test form contains a healthy core of diagnostically informative questions while also highlighting candidates for refinement (e.g., simplifying overly difficult questions or enriching the task design for overly easy ones to restore variability).

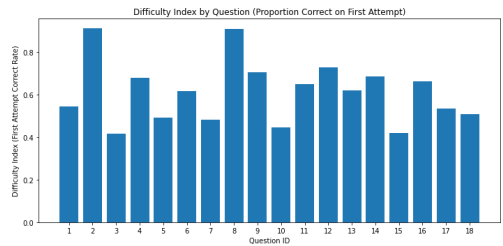

Figure 2: Difficulty index by question (proportion correct on first attempt).

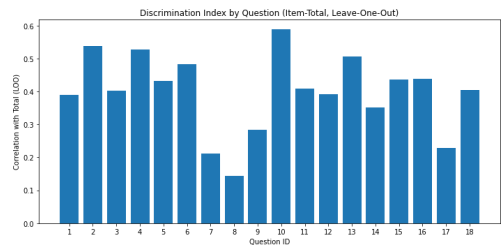

Figure 3: Discrimination index by question (question–total correlation, leave–one–out).

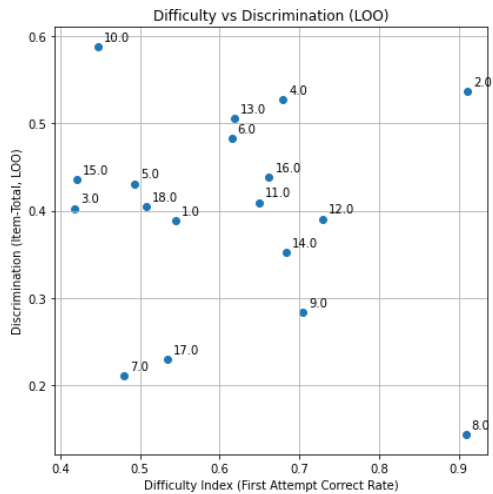

Figure 4: Relationship between difficulty and discrimination. Mid–difficulty questions tend to discriminate best; very easy questions show reduced discrimination.

### 3.3.2 Learning curves by attempt count

To probe within–question learning dynamics, we modeled the cumulative probability of success as a function of attempt number. For each student–question pair, we located the first attempt on which the response was marked correct and computed, for $k = 1, \ldots, 4$, the proportion of pairs solved by or before attempt $k$. To stabilize estimates, questions attempted by fewer than 50 students were excluded from this analysis.

Figure 5 displays the aggregate learning curve across the retained questions. The curve rises sharply from the first attempt (approximately 0.60 ) to the second ( ≈ 0.73 ), indicating that a large share of initially incorrect responses are corrected after a single retry. Gains taper thereafter, reaching ≈ 0.78 and ≈ 0.80 by the third and fourth attempts, respectively, which suggests diminishing returns from repeated opportunities and that most achievable learning occurs early in the iterative cycle.

Question-level trajectories (Figure 6) reveal substantial heterogeneity in both starting level and slope. Questions such as Q2 and Q8 exhibit very high first–attempt correctness (consistent with their difficulty indices near 0.90) and quickly approach mastery, indicating low cognitive load and/or strong alignment with prior knowledge. A second cluster (e.g., Q1, Q3, Q5, Q10) shows marked gains from the first to the second attempt, implying that either the feedback or simple re-engagement successfully bridges initial gaps. By contrast, a subset including Q6, Q15, and Q18 shows flatter post-first-attempt growth; these questions remain well below full mastery by the fourth attempt despite repeated opportunities. This pattern dovetails with earlier findings: Q15, for example, combined relatively low agreement and variable feedback quality, and here likewise exhibits limited convergence, suggesting persistent misconceptions or insufficient scaffolding.

Overall, the attempt-wise learning curves indicate that the platform effectively supports rapid correction on many questions, while also isolating questions that would benefit from targeted instructional revision (e.g., tightening terminology, suppressing answer leakage, and adopting stepwise prompts to foster conceptually grounded revisions rather than

guess–and–check).

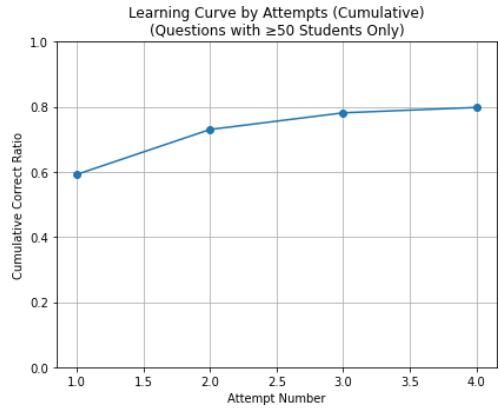

Figure 5: Overall cumulative learning curve across questions with ≥50 students.

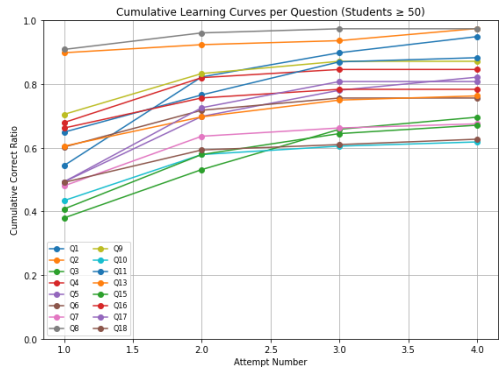

Figure 6: Cumulative learning curves by question (students ≥50). Lines show the proportion solved by or before the indicated attempt.

### 3.3.3 Analysis of feedback impact

Two quantitative measures were introduced to assess the short- and mid-term impact of feedback. First, the *post-feedback correctness conversion rate* was defined as the proportion of initially incorrect responses that were revised to correct answers on the subsequent attempt. This measure captures the immediate instructional effectiveness of feedback. Second, rather than relying solely on first-attempt correctness of subsequent questions, a more targeted *feedback carry-over effect* was examined by focusing on the relationship between prior successful feedback-based corrections and performance on specific target questions. This approach evaluates whether the experience of converting an error into a correct answer through feedback promotes transfer of learning to subsequent, and often more complex, tasks.

To quantify the short-term instructional impact of feedback, we computed the *post-feedback correctness conversion rate*, defined as the proportion of initially incorrect responses that were revised to correct answers on the subsequent attempt following feedback. Across all questions, the conversion rate was 36.7% (=260/708), based on a substantially larger pool of feedback-supported incorrect attempts. This indicates that roughly one in three initially incorrect responses was corrected after a single retry with feedback.

As shown in Table 7, further analysis revealed substantial variation in conversion rates across students, questions, and question types. At the student level, several learners achieved a 100.0% conversion rate with at least three feedback-supported trials (e.g., IDs 2, 17, 26), whereas others remained near zero despite multiple opportunities (e.g., ID 25: 0.0% over 6 trials; ID 51: 7.7% over 26 trials). The median student-level rate among those with ≥3 trials was 40.0%.

At the question level, the highest conversion rates were observed for Q1 (61.5%), Q17 (57.5%), Q11 (47.4%), Q5 (40.7%), and Q4 (38.2%) (threshold: denominator ≥20). By contrast, several problems showed relatively modest gains, including Q7 (21.7%), Q6 (25.0%), Q10 (26.9%), Q18 (27.6%), and Q16 (31.0%).

When grouped by question type (weighted by denominators), closed-ended questions achieved the highest conversion rate at 55.1%, followed by CGT (35.6%) and open-end (33.9%). This pattern suggests that concise, well-defined response formats are more conducive to rapid post-feedback correction, whereas complex open-ended tasks may require additional scaffolding to achieve comparable gains.

Table 7: Post-feedback correctness conversion rates by student group, question, and question type

| Category | Conversion Rate (%) |
|---|---|
| *1. Student-level (representative variation; trials ≥ 3)* | |
| Highest: IDs 2, 17, 26 | **100.0** |
| Median (IQR): | **40.0** (25.0–64.4) |
| Lowest examples: ID 25 (6 trials) | **0.0** |
| ID 51 (26 trials) | **7.7** |
| *2. question-level (Top 5 by rate; denom ≥ 20)* | |
| Q1, Q17, Q11, Q5, Q4 | **61.5, 57.5, 47.4, 40.7, 38.2** |
| *question-level (Bottom 5 by rate; denom ≥ 20)* | |
| Q7, Q6, Q10, Q18, Q16 | **21.7, 25.0, 26.9, 27.6, 31.0** |
| *3. question-type level (weighted by denominators)* | |
| Closed-ended | **55.1** |
| CGT | **35.6** |
| Open-end | **33.9** |

Overall post-feedback conversion: **36.7%** (260/708).

To explore this relationship, we examined three target questions positioned later in the sequence: Q10 (CGT), Q11 (open-ended), and Q18 (CGT). For each target question, students' preceding problem-solving history was analyzed to determine whether they had at least one instance of feedback-driven conversion—defined as an initially incorrect response followed by feedback and an immediate correct answer on the same problem—prior to encountering the target question.

Contingency analysis showed that for Q10, students without prior conversions achieved 10/14 (71.4%) correct, compared to 37/62 (59.7%) among those with such experience ($p$ =

0.547). For Q11, the rates were 11/13 (84.6%) versus 57/64 (89.1%) ($p$ = 0.643). For Q18, students with prior conversions scored 35/55 (63.6%) correct, compared to 2/4 (50.0%) without ($p$ = 0.624).

For first-attempt performance, Q10 showed 24/62 (38.7%) versus 9/14 (64.3%) ($p$ = 0.134); Q11 showed 42/64 (65.6%) versus 8/13 (61.5%) ($p$ = 0.761); and Q18 showed 27/55 (49.1%) versus 2/4 (50.0%) ($p$ = 1.000). Among students who eventually solved the target, Mann–Whitney U tests comparing the attempt number of first success yielded $p$ = 0.123 (Q10), $p$ = 0.897 (Q11), and $p$ = 0.483 (Q18).

None of these comparisons reached statistical significance (all $p$ > 0.05). These results suggest that while feedback-driven conversions were common and often associated with high performance, their direct measurable impact on subsequent complex questions could not be demonstrated in this dataset. The consistently high correctness rates, especially in Q11, likely reflect ceiling effects that mask potential benefits, and several contrasts suffered from small no-prior groups (e.g., Q18: $n$ = 4). The lower first-attempt rate among prior converters in Q10 did not reach significance, which may be due to selection effects or question ordering. Overall, larger samples and a wider distribution of question difficulties are needed to more conclusively assess the long-term transfer of feedback.

### 3.3.4 Analysis of iterative answer refinement within a question

To examine within-question learning, we analyzed iterative answer refinement—sequences where a learner revises an answer across attempts with feedback. Refinement was defined as increasing specificity and conceptual accuracy, rather than parroting model solutions.

For example, learners were asked to justify why the circumcenter is a fair meeting location. A correct explanation requires invoking the equidistance property: the circumcenter is equally distant from all three vertices, ensuring equal travel distance.

Table 8 illustrates a progression from vague to precise reasoning. The initial response ("Equal") lacked clarity. Feedback guided attention to geometric invariance, leading first to a still-ambiguous mention of "length," and ultimately to the correct articulation of equidistance. This trajectory aligns with rubric criteria: clarity increases, feedback is actionable, and the final response employs accurate, domain-specific language.

At scale, such patterns mirror the aggregate results: early gains are strongest when feedback scaffolds concept identification (e.g., "equidistance from vertices") rather than revealing full solutions. This shows how feedback can foster concise, conceptually grounded justifications in geometry tasks.

## 3.4 Learner Self-Assessment and Expert Evaluation

We administered a five-level Likert survey (5 = "strongly agree," 1 = "strongly disagree") and open-ended comments to capture students' perceptions of the activity. The survey addressed five core outcomes: understanding the properties of the circumcenter; representing the circumcenter and circumcircle with a digital tool; explaining how the circumcenter's location varies with triangle shape; reasoning about why the property holds; and recognizing opportunities to apply these ideas in everyday contexts.

Table 8: Iterative Answer Refinement Following Feedback (Student #28; question #6, open-ended: fair meeting point via circumcenter)

| Attempt | Submitted Answer | Judgment (LLM) | Feedback (verbatim) |
|---|---|---|---|
| 1st | "Equal" | INCORRECT | "How does the length of the segment change when you move a vertex of the triangle?" |
| 2nd | "The length of the segment is equal." | INCORRECT | "Think about which property remains unchanged even when moving a vertex of the triangle. Please press the retry button and try again." |
| 3rd | "The distance from the circumcenter to each vertex is equal." | CORRECT | "Correct! The distance from the circumcenter to each vertex of the triangle is always the same." |

Average ratings for these questions ranged from 4.1 to 4.5 on the five-point scale, clustering near the upper end with only modest variation. This pattern indicates broad endorsement across students rather than polarization. A small but non-trivial minority selected "disagree" or "strongly disagree," highlighting areas for improvement in conceptual explanation and tool use.

In the open-ended survey, several recurring themes emerged. On the positive side, students emphasized:

- increased engagement and enjoyment (e.g., "fun," "interesting," "novel");
- the value of hands-on construction and immediate, on-demand feedback (e.g., "I could draw and construct it myself," "got hints right away");
- perceived convenience of the digital tool ("easier," "more convenient"); and
- meaningful connections to real-world contexts (e.g., choosing a fair meeting point with the circumcenter), which many described as motivating and memorable.

A common thread was heightened self-efficacy, as students felt more confident in expressing and applying mathematical knowledge after using the tool and feedback.

Critical remarks were less frequent but consistent. Students mentioned:

- difficulties with software operation (accurate drawing/constructing, occasional errors);
- the need for clearer guidance on constructing the circumcenter and explaining why invariance holds; and
- concerns about feedback, which was at times too vague or too close to the answer.

Taken together, these observations suggest refinements such as (i) short just-in-time micro-tutorials on core constructions with troubleshooting support, (ii) feedback that escalates from conceptual cues to stepwise scaffolds without revealing the full solution, and (iii) optional proof sketches to strengthen conceptual grounding. These adjustments would preserve the high engagement documented here while addressing the usability and explanatory gaps identified by students.

Following the learner survey, we also collected expert evaluations to obtain an external perspective on the system's grading and feedback. Overall, experts judged the automated grading to be broadly reliable across the item set, but they converged on several recurrent failure modes that warrant attention. Chief among these were lexical rigidity and rubric–context mismatches: responses that conveyed the correct mathematical idea were sometimes marked incorrect when key terms were imprecise (e.g., midpoint vs. center, incenter vs. circumcenter, or "two" versus "three" perpendicular bisectors), when the justification referenced an equivalent formulation (e.g., naming the intersection point only implicitly), or when minor, nonessential phrasing was interpreted as conceptually wrong. Reviewers recommended (i) canonicalizing domain synonyms and near-equivalents, (ii) relaxing surface-form rules in favor of concept recognition, and (iii) refining rubrics for open-ended explanations to avoid penalizing mathematically correct but stylistically atypical answers—particularly on items such as Q15, Q3, and Q11, where small terminological slips disproportionately triggered disagreement.

Experts also evaluated the quality of feedback. They were most positive when feedback scaffolded concept discovery without revealing the solution—for example, cueing an invariant ("equal distance from the circumcenter to each vertex"), prompting the relevant construction, and inviting a brief justification. Lower ratings clustered around two problematic patterns: answer leakage (feedback that stated or strongly implied the target statement) and vagueness or tone issues (generic prompts like "last chance" that neither specify the next cognitive move nor sustain a supportive stance). As concrete improvements, reviewers advocated a "concept-first" template (Invariant → Construction → Justification), explicit suppression of solution-revealing phrases, tighter terminology, and tone guidelines that keep prompts actionable, specific, and non-judgmental. Collectively, these refinements would align feedback with expert expectations of formative guidance while preserving learner agency and conceptual precision.

# 4 Conclusion

This study demonstrated the feasibility of integrating large language models (LLMs) with an online geometry exploration platform to deliver automated grading and adaptive feedback for constructive geometry tasks. By embedding LLM-based evaluation into Algeomath and piloting the system with 79 middle-school students, we examined how automated judgments and feedback compared with those of expert teachers in authentic classroom settings. The results showed high overall concordance in grading, with rule-based items yielding perfect agreement and open-ended explanations achieving substantial reliability (overall agreement = 0.866; $\kappa$ = 0.737). Furthermore, short-term learning benefits were evident, as more than one-third of initially incorrect responses were corrected after a single retry with feedback, illustrating the system's capacity to scaffold students in refining misconceptions and completing multi-step geometry tasks. Student surveys confirmed these benefits, reporting increased engagement, heightened self-efficacy, and meaningful connections to real-world contexts. Expert reviews also noted that feedback was most effective when it scaffolded concept discovery without revealing solutions, though concerns remained about lexical rigidity, rubric alignment, and occasional answer leakage.

Despite these promising outcomes, the study also revealed limitations that temper the generalizability of the findings. Current LLMs operate exclusively on textual inputs and therefore cannot directly evaluate the visual precision of geometric constructions, such as the exactness of bisectors or the alignment of intersection points. In addition, the quality and variety of generated feedback remain bounded by the scope of teacher-authored exemplars used in few-shot prompting, which limits adaptability to unanticipated or non-standard student expressions. Statistical analyses further indicated that while short-term corrections were common, transfer to later items was not significant (all $p > 0.05$), a result that may reflect ceiling effects, small group sizes, and the need for more robust longitudinal data.

Looking ahead, several directions for future work are clear. First, incorporating multimodal LLMs capable of processing both text and graphical input would address current limitations in evaluating geometric accuracy. Second, expanding the framework beyond geometry to domains such as algebra and functions would allow a broader assessment of generalizability. Third, leveraging longitudinal student response data could enable more adaptive and personalized feedback through reinforcement learning or retrieval-based methods. Finally, refining feedback design to reduce lexical rigidity, suppress answer leakage, and adopt concept-first scaffolding templates would align automated feedback more closely with expert pedagogical standards.

Taken together, the study underscores both the promise and the challenges of LLM-based formative assessment. With continued refinement, such systems hold the potential to provide scalable, context-aware, and pedagogically aligned support that complements teacher instruction, thereby fostering deeper mathematical engagement and more effective learning experiences for students.

# Acknowledgment

This research was supported in 2024 by the Korea Foundation for the Advancement of Science and Creativity (KOFAC) under the project titled "Development of an Integrated Engineering Tool Model and Prototype for Inquiry-Activity-Based Assessment." The study was reviewed and approved by the Institutional Review Board of Hongik University (IRB No. 7002340-202409-HR021, approval date: September 3, 2024) and conducted in accordance with the principles outlined in the Declaration of Helsinki.